\documentclass[pmlr]{jmlr}

\RequirePackage{graphicx}
 \usepackage{booktabs}
\usepackage{longtable}
\usepackage{algorithm}     
\usepackage{algpseudocode}  
\usepackage{amsmath}  

\makeatletter
\def\set@curr@file#1{\def\@curr@file{#1}} 
\makeatother
\usepackage[load-configurations=version-1]{siunitx} 

\theorembodyfont{\upshape}
\theoremheaderfont{\scshape}
\theorempostheader{:}
\theoremsep{\newline}

\newcommand{\XX}[1]{\textbf{??}}  

\jmlrvolume{298}
\jmlryear{2026}

\title[Laboratory-Specific Sequencing Performance Guarantees in CNV Detection]{Combining Bayesian and Frequentist Inference for Laboratory-Specific Performance Guarantees in Copy Number Variation Detection}

\author{\Name{Austin Talbot}
       \Email{talbota@pillarbiosci.com}\\ 
       \addr Pillar Biosciences Inc\\
       Natick, MA, USA 
       \AND
       \Name{Alex Kotlar}
       \Email{kotlara@pillarbiosci.com}\\ 
       \addr Pillar Biosciences Inc\\
       Natick, MA, USA 
       \AND
       \Name{Yue Ke}
       \Email{key@pillarbiosci.com}\\ 
       \addr Pillar Biosciences Inc\\
       Natick, MA, USA 
       }

\begin{document}

\maketitle

\begin{abstract}
Targeted amplicon panels are widely used in oncology diagnostics, but providing per-gene performance guarantees for copy number variant (CNV) detection remains challenging due to amplification artifacts, process-mismatch heterogeneity, and limited validation sample sizes. While Bayesian CNV callers naturally quantify per-sample uncertainty, translating this into the frequentist population-level guarantees required for clinical validation, coverage rates, false-positive bounds, and minimum detectable copy-number changes, is a fundamentally different inferential problem. We show empirically that even robust Bayesian credible intervals, including coarsened posteriors and sandwich-adjusted intervals, are severely miscalibrated on panels with small amplicon counts per gene. To address this, we propose a hybrid framework that evaluates Bayesian posterior functionals on validation samples and models the resulting squared losses with a Gamma distribution, yielding tolerance intervals with valid frequentist coverage. Three components make the method practical under real-world constraints: (1) imputation that removes the influence of true CNV-positive samples without requiring known ground truth, (2) regularization to address small sample variability, and (3) evidence-based stratification on the log model evidence to accommodate non-exchangeable noise profiles arising from process mismatch. Evaluated on two targeted amplicon panels using leave-one-out cross-validation, the proposed method achieves single-digit mean absolute coverage error across all genes under both process-matched and unmatched conditions, whereas Bayesian comparators exhibit mean absolute errors exceeding 60\% on clinically relevant genes such as ERBB2.
\end{abstract}

\section{Introduction}\label{sec1}

A false-positive ERBB2 amplification call can route a breast cancer patient onto trastuzumab, a targeted therapy that carries a 4.0\% rate of cardiac events at seven-year follow-up even in patients with confirmed HER2-positive disease~\citep{romond2012seven}, all for a molecular subtype she does not carry~\citep{fehrenbacher2020nsabp}. In colorectal cancer, a missed KRAS mutation can lead to cetuximab treatment that offers no survival benefit whatsoever (HR 0.98, $P = .89$) while exposing the patient to dermatologic and infusion-related toxicity~\citep{karapetis2008k}. In both cases the analytical error is upstream: the sequencing-based diagnostic that informed the treatment decision was either insufficiently sensitive or insufficiently specific. The clinician had no way to know this, because the test report carried generic performance guarantees based on internal data, as opposed to ``for this panel, in this run, the false-positive rate for ERBB2 copy-number calling is below $x$\% with $y$\% confidence.'' \textbf{Providing exactly that kind of individualized guarantee is the purpose of this work.}

The targeted amplicon panels this method was developed for offer considerable advantages compared to whole genome sequencing: quick turnaround times, compatibility with low DNA content (critical for later-stage patients who cannot donate much blood), tolerance of substantial sample degradation, and cost-effectiveness~\citep{sie2014performance,gao2019next}. The ability to produce results under these demanding conditions has proven invaluable to thousands of patients. However, this same setting creates a challenging environment for analysis, particularly for CNV detection, where limited amplicons~\citep{boeva2014multi}, process-matching difficulties~\citep{derouault2020covcopcan}, and amplification artifacts~\citep{peng2015reducing} all degrade analytical performance. 

The CNV caller we use, BayesCNV~\citep{talbot2026bayescnv}, is fundamentally Bayesian. Bayesian methods are a natural fit for diagnostics~\citep{bours2021bayes}. They naturally incorporate prior information~\citep{lee2024using}, quantify uncertainty~\citep{gelman2020bayesian}, and evaluate model evidence~\citep{lotfi2022bayesian}, the last of which serves as an indicator of questionable samples~\citep{talbot2026detecting}. At the individual level, these qualities are irreplaceable.

However, providing per-gene performance guarantees for a given laboratory is a fundamentally frequentist question: it asks what sensitivity, false-positive rate, or estimation error a clinician can expect across future samples processed on that platform. Bayesian credible intervals can exhibit poor frequentist coverage under model misspecification — the posterior concentrates with a covariance differing from the frequentist sandwich variance~\citep{kleijn2012bernstein}, yielding strictly higher risk~\citep{muller2013risk}, and misspecification can even render the posterior inconsistent~\citep{grunwald2017inconsistency}. In the amplicon setting, such misspecification is a practical certainty, as no parametric model fully captures the target-specific biases of PCR amplification.

Fortunately, clinical validation protocols require running multiple samples in an initial run, providing genuine repeated observations from which we can evaluate any posterior functional of interest to obtain an empirical sampling distribution with valid frequentist coverage regardless of model correctness. Due to difficulties in procuring samples with known CNV status, we instead use the discrepancy between the posterior mean and the CNV-neutral value of 1 to evaluate variability. We develop a novel method to reduce the influence of true positives via conditional imputation of the largest CNVs in log space, then model the resulting losses with a gamma distribution,  a technique adapted from actuarial practice. Given the limited number of samples typically available, we also implement a conjugate prior composed of pseudo-observations based on previous laboratory results. The resulting procedure is robust to a subset of samples containing true positives, computationally tractable, reliable under small sample sizes, and naturally accommodates prior information from previous experiments.

To better characterize performance when samples are not process-matched, we augment our method to stratify samples based on how well they align with model assumptions, quantified via the Bayesian evidence. Prior work has shown that the Bayesian evidence serves as an indicator of process mismatch or low sample quality~\citep{talbot2025classifying,talbot2026detecting}. When the sample size is sufficiently large, we can split the samples into two groups based on evidence score and compute tolerance intervals using the above method on each group. 

We evaluate our method across two targeted amplicon panels, using a leave-one-out scheme to assess calibration of our tolerance intervals. We show empirically (1) that Bayesian credible intervals (even with robust inference) have inadequate frequentist properties, (2) that our method provides proper coverage, even in noisy data, and (3) that stratification and imputation are critical when samples are not process matched and contain positive samples respectively. 

The remainder of this work is organized as follows. Section~\ref{sec2} contains work that is either related to or inspired this work. Section~\ref{sec3} details our method, which is tested on synthetic data in Section~\ref{sec4} and observed data in Section~\ref{sec5}. Finally, in Section~\ref{sec6} we provide brief remarks and directions for future work.

\subsection*{Generalizable Insights about Machine Learning in the Context of Healthcare}
\begin{itemize}

\item \textbf{Robust Bayesian methods are insufficient for frequentist guarantees.} Even with robustified posteriors, hard-to-model noise sources can prevent reliable achievement of the nominal coverage rates required for clinical use.
\item \textbf{Non-IID noise profiles can make error bounds inaccurate.} Methods that assume all samples have similar noise profiles can provide misleading guarantees when this assumption is violated.
\item \textbf{Bayesian and frequentist methods are complementary.} Bayesian methods excel at the individual level, allowing the incorporation of prior information and evaluation of model assumptions. Frequentist methods are a natural choice for population-level performance guarantees needed for product validation.  
\end{itemize}

\section{Related Work}\label{sec2}
Read-depth-based CNV detection from targeted sequencing spans several methodological families. Normalization-then-segmentation pipelines — ONCOCNV~\cite{boeva2014multi}, CNVkit~\cite{talevich2016cnvkit}, CONTRA~\cite{li2012contra} — correct for GC content and library-size biases before applying CBS or threshold-based calling. PCA/SVD denoising methods such as XHMM~\cite{fromer2012discovery} and CoNIFER~\cite{krumm2012copy} remove batch effects from the coverage matrix. Probabilistic count models jointly estimate artifacts and copy-number states: CODEX2~\cite{jiang2018codex2} via Poisson latent factors and GATK-gCNV~\cite{babadi2023gatk} via a negative-binomial hierarchical HMM. Mixture-of-Poissons methods, cn.MOPS~\cite{klambauer2012cn} and panelcn.MOPS~\cite{povysil2017panelcn}, model cross-sample variation as discrete copy-number components, while ExomeDepth~\cite{plagnol2012robust} and DECoN~\cite{fowler2016accurate} use beta-binomial likelihoods with optimized reference sets. Benchmarking studies consistently report dataset- and laboratory-dependent performance~\cite{moreno2020evaluation,munte2025detection}, but none of these tools provide laboratory-customized frequentist performance guarantees.

An alternative strategy is to build robustness directly into the modeling stage. Power posteriors raise the likelihood to an exponent $\tau < 1$ to reduce sensitivity to misspecification~\cite{miller2019robust, holmes2017assigning}, and generalized Bayesian inference replaces the log-likelihood with an arbitrary loss, with learning rates calibrated via the SafeBayes framework~\cite{bissiri2016general, grunwald2017inconsistency}. These approaches improve per-sample posterior inference but address a fundamentally different problem: they seek a better posterior for a single observation under a misspecified model, whereas we require frequentist guarantees — coverage, sensitivity, mean squared error — that hold across a population of future samples.

\section{Methods}\label{sec3}

\subsection{Background and Statistical CNV Model}

We consider $N$ samples assayed on a targeted amplicon panel spanning $J$
genes, where gene~$j$ is tiled by $n_j$ amplicons ($j=1,\dots,J$).  For a
given sample let $s_{j,k}$ and $r_{j,k}$ denote the read counts at
amplicon~$k$ of gene~$j$ for the test and diploid reference samples,
respectively.  Because the model is fit independently to each sample, we
suppress the sample index throughout this subsection.

\subsubsection{Features}
For each amplicon we form a log copy-number ratio (lCNR) relative to the
normal baseline.  Amplicon-specific capture biases cancel in the ratio,
isolating the copy-number signal.  The raw lCNR is
\begin{equation}\label{eq:raw_lcnr}
  \tilde{X}_{j,k}
    \;=\; \log\!\bigl(s_{j,k} + c\bigr)
        - \log\!\bigl(r_{j,k} + c\bigr),
\end{equation}
where $c>0$ is a small pseudo-count guarding against zero counts.  We remove
global depth differences by median-normalizing across all amplicons:
\begin{equation}\label{eq:norm_lcnr}
  X_{j,k}
    \;=\; \tilde{X}_{j,k}
        - \operatorname{median}_{j',k'}\!\bigl(\tilde{X}_{j',k'}\bigr).
\end{equation}
After normalization, $X_{j,k}=0$ indicates a copy-neutral amplicon, negative
values indicate deletions, and positive values indicate amplifications.  The
median is robust to the sparse copy-number alterations expected in any single
sample.

\subsubsection{Statistical Model}
We model the normalized lCNRs with the hierarchical Bayesian framework
\textsc{BayesCNV}~\citep{talbot2026bayescnv}.  Gene-level means $\mu_j$ are
partially pooled toward a global mean $\mu_0$ through a between-gene variance
$\sigma^2$.  Amplicon-level noise is factored into a global scale $\tau_0^2$
and gene-specific multipliers $z_j^2$, giving per-gene scales
$\tau_j^2 = \tau_0^2 z_j^2$.  The generative model is:
\begin{equation}\label{eq:bayescnv}
\begin{aligned}
  \mu_0 &\sim \mathcal{N}\!\bigl(0,\;10^2\bigr), &
  \sigma^2 &\sim \operatorname{Inv\text{-}Gamma}(\alpha_\sigma,\;\beta_\sigma), \\
  \mu_j \mid \mu_0,\sigma^2
    &\sim \mathcal{N}\!\bigl(\mu_0,\;\sigma^2\bigr), &
    &j=1,\dots,J, \\[4pt]
  \tau_0^2
    &\sim \operatorname{Inv\text{-}Gamma}(\alpha_{\tau_0},\;\beta_{\tau_0}), &
  z_j^2
    &\sim \operatorname{Inv\text{-}Gamma}(\alpha_\tau,\;\beta_\tau), \\
  \tau_j^2 &= \tau_0^2 z_j^2, \\[4pt]
  X_{j,k} \mid \mu_j,\tau_j
    &\sim \operatorname{SoftLaplace}\!\bigl(\mu_j,\;\tau_j\bigr), &
    &k=1,\dots,n_j,
\end{aligned}
\end{equation}
where $\operatorname{SoftLaplace}$~\citep{bingham2019pyro} is a heavy-tailed
location--scale family providing robustness to outlier amplicons.  Throughout,
$\mathcal{N}(\mu,\,\sigma^2)$ denotes a Normal distribution with mean~$\mu$
and \emph{variance}~$\sigma^2$.  The inverse-gamma hyperparameters are set to
mildly regularizing values that keep the posterior proper while exerting
minimal influence in data-rich genes.

\subsubsection{Inference}
The inferential target is the gene-level mean~$\mu_j$: values near zero
indicate diploid copy number; substantial deviations signal deletions or
amplifications.  Posterior samples are drawn via Hamiltonian Monte
Carlo~\citep{betancourt2017conceptual} from
$p(\theta\mid\mathbf{X})$, with
$\theta = \bigl\{\mu_0,\sigma^2,\{\mu_j,z_j^2\}_{j=1}^{J},\tau_0^2\bigr\}$.
From each marginal posterior of~$\mu_j$ we form $(1-\gamma)$
highest-posterior-density credible intervals and flag gene~$j$ as altered
whenever the interval excludes zero.  We also estimate the log model evidence
$\mathcal{Z} = \log \int p(\mathbf{X}\mid\theta)\,p(\theta)\,\mathrm{d}\theta$
via thermodynamic integration~\citep{gelman1998simulating}, which serves
both as a model-comparison criterion and, in
Section~\ref{sec:anti-tempered}, as a process-matching diagnostic.
Further details appear in~\citet{talbot2026bayescnv}.

\subsection{Removing Positives with Conditional Order Imputation}\label{ssec:impute}

In a clinical deployment the caller is validated on $K$ samples processed by
the client laboratory.  Ideally one would evaluate sensitivity and specificity
against known positive and negative controls, but obtaining such controls at
scale is rarely feasible.  We therefore adopt a strategy that requires no
prior knowledge of which samples carry copy-number alterations, exploiting
the sparsity of CNV events: across the $J$~genes of the panel, the vast
majority of gene--sample pairs are diploid.

For any gene~$j$, the $K$ posterior means
$\hat\mu_j^{(1)},\ldots,\hat\mu_j^{(K)}$ are dominated by a diploid
component near zero, with at most $m$ samples harboring a genuine
amplification.  The top-$m$ values---those most likely to reflect true gains---are
the primary source of contamination in a tolerance limit estimated from the
empirical distribution.  We impute these suspect values from the bulk
(CNV-neutral) distribution, constructing a pseudo-diploid reference from which
calibrated tolerance limits can be derived.  This correction is deliberately
one-sided: copy-number gains can be arbitrarily large, inflating the upper
tail, whereas due to low tumor purity deletions, even homozygous losses, depress
$\hat\mu_j^{(s)}$ only modestly.

\begin{figure}[ht]
  \centering
  \includegraphics[width=\textwidth]{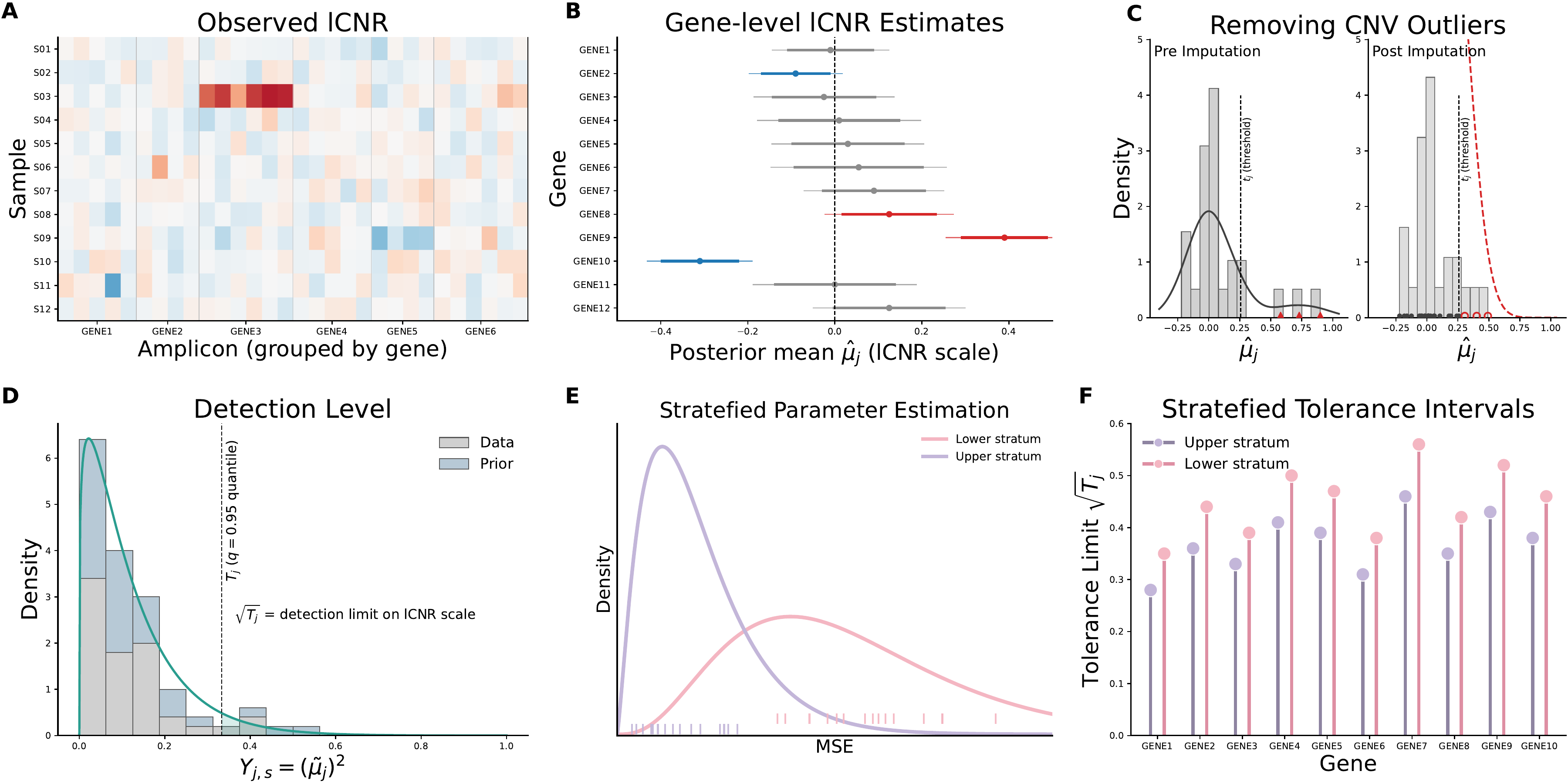}
  \caption{%
\textbf{(A)} Heatmap of normalized lCNRs $X_{j,k}$ for representative samples (rows) across amplicons grouped by gene (columns). 
\textbf{(B)} Forest plot of BayesCNV posterior means $\hat{\mu}_j$ with $(1{-}\gamma)$ HPD credible intervals for one representative sample.
\textbf{(C)} Conditional order-statistic imputation for one representative gene. \textit{Left:} distribution of $K$ posterior means with top-$m$ suspect values (filled red markers) and threshold $t_j$ (dashed line). \textit{Right:} distribution after replacing suspect values with order-statistic draws.
\textbf{(D)} Gamma model fit (solid colored curve) to the empirical distribution of squared imputed posterior means $Y_{j,s}$ (grey) and pseudo observations (pink); the vertical dashed line marks the tolerance quantile $T_j$, and $\sqrt{T_j}$ is the minimum detectable lCNR deviation.
\textbf{(E)} Stratefied parameter estimates each sample group divided by model evidence.
\textbf{(F)}  Per-gene tolerance limits $\sqrt{T_j}$ in upper and lower evidence strata. 
  }
  \label{fig:exceedance_sweep}
\end{figure}


Fix gene~$j$ and write the $K$ ordered posterior means as
\begin{equation}\label{eq:ordered_mu}
  \hat\mu_{j,(1)}
    \;\leq\; \hat\mu_{j,(2)}
    \;\leq\; \cdots
    \;\leq\; \hat\mu_{j,(K)}.
\end{equation}
Fix a small integer $m < K$, set $L = K - m$, and define the threshold
\begin{equation}\label{eq:threshold}
  t_j \;=\; \hat\mu_{j,(L)},
\end{equation}
the largest value among the $L$ retained observations.  The top-$m$
values $\hat\mu_{j,(L+1)},\ldots,\hat\mu_{j,(K)}$ are replaced by the
following procedure, applied independently per gene.

\begin{enumerate}
\item \textbf{Bulk model fit.}
  Fit a Normal distribution
  $\hat F_j = \mathcal{N}\!\bigl(\hat\xi_j,\,\hat\omega_j^2\bigr)$
  to the $L$ retained values using robust estimators:
  \begin{equation}\label{eq:robust_fit}
    \hat\xi_j
      = \operatorname{median}\!\bigl(
        \hat\mu_{j,(1)},\ldots,\hat\mu_{j,(L)}
      \bigr),
    \qquad
    \hat\omega_j
      = 1.4826\;\operatorname{MAD}\!\bigl(
        \hat\mu_{j,(1)},\ldots,\hat\mu_{j,(L)}
      \bigr),
  \end{equation}
  where $\operatorname{MAD}$ is the median absolute deviation and
  $1.4826 = 1/\Phi^{-1}(3/4)$ is the Fisher consistency factor for
  Gaussian scale estimation
  (see Appendix~\ref{app:derivations};
  \citealp{HuberRonchetti2009RobustStatistics}).

\item \textbf{Truncated sampling.}
  Let $S_j = 1 - \hat F_j(t_j)$.  Draw
  $U_1,\ldots,U_m \overset{\mathrm{i.i.d.}}{\sim}
  \operatorname{Uniform}(0,1)$ and set
  \begin{equation}\label{eq:trunc_draw}
    Z_i
      =
      \hat F_j^{\,-1}\!\Bigl(
        \hat F_j(t_j) + U_i \, S_j
      \Bigr),
      \qquad i = 1,\ldots,m.
  \end{equation}
  Each $Z_i$ is a draw from $\hat F_j$ truncated to
  $[t_j,\infty)$.

\item \textbf{Order-statistic replacement.}
  Sort the draws to obtain $Z_{(1)} \leq \cdots \leq Z_{(m)}$ and set
  \begin{equation}\label{eq:replace}
    \tilde\mu_{j,(L+i)} = Z_{(i)},\; i = 1,\ldots,m;
    \qquad
    \tilde\mu_{j,(r)} = \hat\mu_{j,(r)},\; r = 1,\ldots,L.
  \end{equation}
\end{enumerate}

\noindent
This procedure reproduces the classical conditional distribution of
upper order statistics given $X_{(L)} = t$ for an i.i.d.\ continuous
sample~\citep{david2004order}, with the true $F$ replaced by the robust
estimate~$\hat F_j$.

\subsection{Tolerance Intervals Via a Gamma Quantile}
\label{sec:eb_tolerance}

Having removed the influence of up to $L$ positive samples using the method in Section~\ref{ssec:impute}, we now work to obtain tolerance intervals on the posterior means by gene. Note, to maintain uncertainty in the imputations, this procedure will be repeated multiple times with different imputed values and use the empirical mean estimate. However, for clarity, we will suppress this in the following sections.

We characterize the CNV performance via the MSEs of the posterior means (assuming the sample is CNV neutral). That is 
\begin{equation}\label{eq:sq_lcnr}
  Y_{j,s}
    \;=\;
    \bigl(\tilde\mu_j^{(s)}\bigr)^{\!2}.
\end{equation}
A natural model for $Y$ is a Gamma distribution. If $\tilde\mu_j^{(s)} \sim \mathcal{N}(0,\,\tau_j^2)$ then
$Y_{j,s} \sim \tau_j^2\,\chi^2_1 =
\operatorname{Gamma}\!\bigl(\tfrac{1}{2},\,2\tau_j^2\bigr)$.
More generally, a residual process-mismatch bias $\delta_j$ gives
\begin{equation}\label{eq:ncx2}
  \tilde\mu_j^{(s)} \sim \mathcal{N}\!\bigl(\delta_j,\,\tau_j^2\bigr)
  \quad\implies\quad
  Y_{j,s}
    = \tau_j^2\,\chi^2_1(\lambda_j),
  \qquad
  \lambda_j = \delta_j^2/\tau_j^2,
\end{equation}
a scaled noncentral chi-squared variate~\citep{urkowitz2005energy}.
Moment-matching to $\operatorname{Gamma}(\alpha_j,s_j)$ yields
(see Appendix~\ref{app:derivations} for the derivation)
\begin{equation}\label{eq:gamma_moments}
  \alpha_j
    =
    \frac{(1+\lambda_j)^2}{2(1+2\lambda_j)},
  \qquad
  s_j
    =
    \frac{2\tau_j^2(1+2\lambda_j)}{1+\lambda_j}.
\end{equation}
When $\delta_j=0$ we recover $a_j=\tfrac{1}{2}$; a nonzero bias
inflates $a_j$ in proportion to $\lambda_j$, so a flexible Gamma
MLE accommodates both regimes.  Additional motivation for the Gamma
family is given in Appendix~\ref{app:derivations}.

Once we estimate the parameters of the gamma distribution, we can construct a $1-p$ tolerance interval as the quantile $T_j=F^{-1}(\hat{\alpha}_j,\hat{s}_j)$. This can be converted to a tolerance interval on the original CNV estimate as the square root (loss is just deviation squared). Given that this transformation is monotonic it inherits coverage properties. As such, our guarantee is that a CNV-neutral sample with probability $1-p$ will have a CNV estimate in the range
\begin{equation}
Pr(\hat{\mu}_j^{new}\notin[-\sqrt{T_j},\sqrt{T_j}])<p.
\end{equation}
Note that while the interval is valid, if $\delta_j\ne0$ the coverage will be lopsided.

While this is framed as a bound on the expected variability of CNV estimates, we also can use this as a proxy for expected performance for positive samples as well. Provided that the amplicon-level biases and variability is independent of CNV status, this is roughly the certainty we would see in lCNR estimates of positive samples as well. In particular, $\exp(\sqrt{T_j})$ is the lowest detectable CNV for gene $j$.

\subsection{Incorporating a Conjugate Prior with Pseudo-Observations}
\label{sec:pseudo}

Quantile estimation can be inherently noisy, particularly when in the upper quantiles like those used in this work. This is a particularly acute issue in our applications as customers often repeatedly use different combinations of samples, both to evaluate robustness and also search for combinations that yield the ``best'' results. This can result in a particularly difficult combination when this is run at low sample sizes ($N=10$). Our applications require low variance, even at the expense of a higher MSE due to increased bias. 

We achieve this by placing a prior on both $\alpha_j$ and $s_j$. We can create a conjugate prior for any regular model where $A=\{x:p(\theta|x)>0\}$ does not depend on $\theta$, that is support does not depend on the parameters, by the incorporation of pseudo observations $(\xi_1,\dots,\xi_{N'})$~\citep{bickel2015mathematical}. That is, our prior is 
$p(\alpha_j,s_j)=\prod_{i=1}^{K'} p(\xi_k|\alpha_j,s_j)\Bigl/\int \prod_{k=1}^{N'} p(\xi_k|\alpha_j,s_j)$
resulting in a posterior of 
\begin{equation}
    p(\alpha_j,s_j|Y_{j,s},\{\xi\}) = \prod_{k=1}^{K'+K} p(\xi'_i|\alpha_j,s_j)\Bigl/\int \prod_{k=1}^{K'+K} p(\xi'_k|\alpha_j,s_j),
\end{equation}
where $\{\xi_k'\}_{k=1:K+K'}=\{Y_{j,1}\dots,Y_{j,s},\xi_1,\dots,\xi_{K'}\}$. The advantage of this posterior is that computation is straightforward with MAP estimation, simply augment the observed data with pseudo-observations and perform maximum likelihood.

While the easiest method for the above prior is to simply tack on a small number of  samples, a less noisy method is to incorporate a large number of samples and weight their contribution in the likelihood to achieve a fixed sample size. We propose two methods for choosing these prior samples. The first is to use samples from a previous related experiment. This method has the downside that it either (i) requires the customer to run a first experiment without a prior or (ii) involves using one customer's data for a separate group. The second is to use an internal run to estimate $\alpha_j$ and $s_j$ under standard conditions, then generate a large number of pseudo observations with those parameters. This is the method we have chosen commercially.

\subsection{Evidence-Stratified Tolerance Limits}
\label{sec:anti-tempered}

The tolerance limits in Section~\ref{sec:eb_tolerance} assume that the $K$
 samples are exchangeable draws from a single production
population.  We frequently find that not all samples are process matched (i.e.\
sequenced with the same instrument, library preparation, and analysis
pipeline used clinically), introducing bias and excess variance
(Appendix~\ref{app:process_matching}).  When samples are pooled, the squared posterior means $Y_{j,s}$
follow a mixture of Gamma-like distributions whose components may differ
in both location and scale due to different noise profiles.  Fitting a single
$\operatorname{Gamma}(\alpha_j,s_j)$ to such a mixture can yield
tolerance quantiles that differ dramatically from the true value of each group individually. 

We eliminate this artifact by stratifying the validation samples on the
log model evidence
$Z_s = \log p(\mathbf{X}_s \mid \mathcal{M})$, which serves as a
process-matching diagnostic~\citep{talbot2026detecting}: well-matched
samples yield high evidence while mismatched samples yield low evidence. There are a variety of techniques for approximating this value, such as thermodynamic integration~\citep{gelman1998simulating} or annealed importance sampling~\citep{neal2001annealed}, or in this work a Laplace approximation~\citep{gelman1995bayesian}.
Partitioning on $\mathcal{Z}_s$ produces strata within which the
single-Gamma model is approximately correctly specified.

Concretely, let $Z_{\mathrm{med}}$ denote the sample median of
$\{Z_1,\ldots,Z_K\}$.  We define two strata:
\begin{equation}\label{eq:strata}
  \mathcal{S}^{+}
    = \bigl\{s : Z_s \geq Z_{\mathrm{med}}\bigr\},
  \qquad
  \mathcal{S}^{-}
    = \bigl\{s : Z_s < Z_{\mathrm{med}}\bigr\},
\end{equation}
with $|\mathcal{S}^{+}| = \lceil K/2 \rceil$ and
$|\mathcal{S}^{-}| = \lfloor K/2 \rfloor$.  Within each stratum
$g \in \{+,-\}$ and for each gene~$j$, we fit a separate Gamma model to
the squared imputed posterior means
$\{Y_{j,s} : s \in \mathcal{S}^{g}\}$, augmented with the
pseudo-observations described in
Section~\ref{sec:pseudo}, yielding per-stratum tolerance quantiles
\begin{equation}\label{eq:strat_tol}
  T_j^{(g)}
    = F^{-1}_{\operatorname{Gamma}(\hat\alpha_j^{(g)},\,
      \hat s_j^{(g)})}\!\bigl(1-p\bigr),
  \qquad g \in \{+,-\}.
\end{equation}
The reported tolerance limit for gene~$j$ in each sample would then be $T_{j,s}=T_j^{g(s)}$.

This procedure requires $K > 20$ so that each stratum retains at least
ten observations, which in our experience is the number required to accurately estimate the parameters of the gamma distribution. This stratification is also optional, one potential method is to detect whether there is evidence for non-exchangability in the samples via a test like the one developed in~\citet{talbot2026detecting}, allowing for formal assessment of whether such stratification is even necessary.

\section{Synthetic Results}\label{sec4}
The purpose of this synthetic section is to evaluate the impact that the imputation scheme and prior have on the quantile estimates. We evaluate the bias, variance, and mean squared error method as a function of sample size, with a particular focus on the
contributions of the prior distribution and the outlier-imputation scheme.
Distribution parameters $\delta = 0.2$ and $\tau^2 = 0.17$ were estimated
from a held-out panel analyzed in Section~\ref{ssec:hs341}; importantly, these
values are drawn from a different panel than the one used to construct
the informative prior, ensuring that the prior is not fit to the evaluation
data. Synthetic datasets were repeatedly drawn at sample sizes ranging from
$N = 5$ to $N = 50$ from $\mathcal{N}(\delta, \tau^2)$, and for each draw we
computed the true quantile bounding $|X|$, comparing four estimators. First the standard maximum-likelihood estimate (Empirical), used as a baseline. Second, our novel estimator with no prior information and $m=1$. Third, the same method but with the informative prior and an effective sample size of $5$. And Finally, our estimator with the same 
          informative prior and an imputation weight that scales linearly with
          $N$, representing the regime in which the number of outliers grows
          proportionally with the number of samples processed. The true value was $0.665$. 

\begin{figure}[ht]
  \centering
  \includegraphics[width=\textwidth]{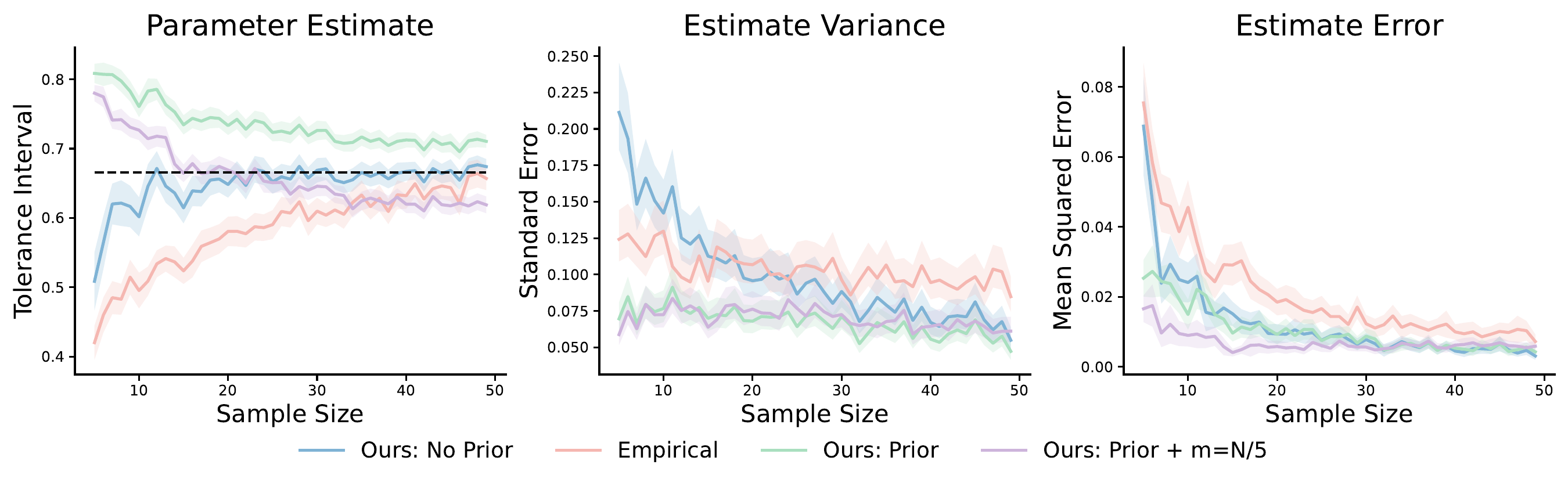}
  \caption{Interval estimation performance as a function of sample size for
    the four estimators.
    \textbf{Left:} posterior mean parameter estimate with $95\%$ confidence
    intervals. \textbf{Center:} standard error of each estimator.
    \textbf{Right:} mean squared error. Data were generated from
    $\mathcal{N}(\delta, \tau^2)$ with $\delta = 0.2$ and $\tau^2 = 0.17$
    over $N \in [5, 50]$.}
  \label{fig:synthetic1}
\end{figure}

Results are displayed in Figure~\ref{fig:synthetic1}. Relative to the
empirical baseline, the proposed estimator without a prior reduces bias and MSE
at the cost of increased variance; at small sample sizes the standard error
reaches approximately $25\%$ of the true parameter value, a level that is
prohibitively high for reliable clinical decision-making. Incorporating the
informative prior yields a substantial reduction in variance, roughly
$2.5\times$, bringing the relative standard error to approximately $10\%$, a
range consistent with stable performance guarantees under the variability
introduced by routine sample-selection interventions. Finally, the
$m = 0.2N$ condition demonstrates that the imputation scheme remains
well-calibrated even when the outlier proportion grows with sample size:
neither bias nor MSE increases materially relative to the fixed-$m$ setting,
confirming that the approach is robust to elevated rates of true-positive
contamination.

\section{Results} \label{sec5}
We evaluate the empirical coverage of the proposed tolerance intervals against two
alternatives drawn from the Bayesian robustness literature, along with traditional MSE-based coverage estimate corresponding to a $\text{Gamma}(1/2,s)$ distribution.  All methods are assessed
on a common criterion: for each gene~$j$ and each diploid validation sample~$s$, we
record whether the interval excludes zero, and report the resulting false-positive rate
over $(j, s)$ pairs relative to the nominal level~$\gamma$.  The first comparator is
the standard $(1-\gamma)$ highest-posterior-density (HPD) credible interval for
$\mu_j^{\text{new}}$ obtained directly from \textsc{BayesCNV}.  The Bernstein--von
Mises theorem guarantees that this interval has asymptotically correct frequentist
coverage as the number of amplicons per gene $n_j\to\infty$; however, for the targeted
amplicon panels considered here, $n_j$ is small (typically $4$--$20$), placing the
data squarely outside this asymptotic regime. The remaining three comparators apply misspecification-robust modifications to the
\textsc{BayesCNV} posterior for each new sample and are the  coarsened posterior of
\citet{miller2019robust}, and the sandwich posterior of \citet{muller2013risk}.

We evaluate coverage using leave-one-out cross-validation: for each
validation sample~$s$, the tolerance interval is estimated from the remaining $K-1$
individuals and we record whether $|\hat{\mu}_j^{(s)}| \leq \sqrt{T_j}$.  Empirical
coverage is then reported alongside interval width for each method, with width serving
as the secondary criterion, narrower intervals that maintain nominal coverage
correspond directly to a lower minimum detectable CNV~$\exp(\!\sqrt{T_j})$ and hence
greater clinical sensitivity.

\subsection{Small Panel LBX}

We first analyze a dataset comprising of 32 FFPE samples sequenced with a 172-amplicon panel
designed for low-cost cancer screening (targeting five CNV-associated genes).
Of these samples, 10 libraries were prepared from
substantially fragmented FFPE DNA, and 22 from higher-quality clinical FFPE
material. We generated data as follows. We chose 5 of the 32 samples to create an averaged normal reference, ranging from 0 samples being fragmented DNA to all 5. We then evaluated the caller on the remaining 22 samples. This was repeated 30 times with different normal selection, yielding a total of $6\cdot30\cdot 22$ different runs. As such, we carefully control the level of known process-mismatch in the samples. For each sample, we also evaluated the marginal likelihood using a Laplace approximation.

\subsubsection{Process Matched Coverage}
\label{ssec:essential_matched}

We quantify calibration via the \emph{mean absolute coverage error} (MACE),
\begin{equation}
  \mathrm{MACE} = \frac{1}{|\Gamma|}\sum_{\gamma\in\Gamma}\bigl|\hat{C}(\gamma)-\gamma\bigr|,
  \label{eq:mace}
\end{equation}
where $\Gamma$ is a uniform grid of nominal coverage levels on $[0.7,1]$ and
$\hat{C}(\gamma)$ is the empirical coverage at level $\gamma$ evaluated per gene
over leave-one-out folds; values are reported multiplied by $100$
(Table~\ref{tab:essential_panel_coverage_mae}).
Under matched conditions all three \textsc{BayesCNV}-based comparators are
severely miscalibrated. The standard HPD and sandwich \citep{muller2013risk}
intervals are drastically overconfident---their MACE values, reaching up to
$66$ and $69$ respectively, reflect intervals so narrow that they fail to
contain the true signal across most of the $[0.7,1]$ coverage range; this is
the expected consequence of applying a large-sample posterior approximation at
the small amplicon counts ($n_j\approx4$--$20$) of this panel.
The coarsened posterior \citep{miller2019robust} is erratic: near-zero MACE
on some genes indicates the opposite failure mode---intervals so wide that
empirical coverage is $100\%$ throughout---while MACE as high as $42$ on other
genes indicates severe overconfidence, with both failure modes present
simultaneously depending on the gene.
The proposed tolerance intervals and the MSE-based estimator together form a
clearly distinct tier, each achieving single-digit MACE across all five genes
and representing a substantial improvement over all three Bayesian comparators.
Within this tier, the proposed intervals are the best or tied-best on every
gene, with the largest margin on ERBB2 ($3.0\pm0.2$ vs.\ $5.6\pm0.5$ for MSE),
a gene of direct clinical relevance for copy-number calling.

\subsubsection{Unmatched Coverage}
\label{ssec:essential_unmatched}

Under maximum process mismatch the failure modes of the Bayesian comparators
become more varied and in some respects more severe, as shown in
Table~\ref{tab:essential_panel_coverage_mae} and
Figure~\ref{fig:exceedance_sweep}.
Process mismatch inflates the posterior variance on several genes, causing
the standard, coarsened, and sandwich intervals to swing from overconfidence
to near-total underconfidence (empirical coverage approaching $100\%$, MACE
near zero) on genes such as KIT and PIK3CA---a change in failure direction, not
a genuine improvement in calibration.
On ERBB2, where mismatch does not rescue the posterior variance, all three
Bayesian variants remain severely overconfident (MACE$\,{\times}\,100$ up to
$74$).
This instability, intervals that are either far too wide or far too narrow
depending on the gene and the degree of process mismatch, renders the Bayesian
comparators unreliable for clinical use.
The proposed tolerance intervals remain uniformly well-calibrated across all
five genes, as does the MSE-based estimator, which degrades modestly but
remains in the single-digit MACE regime.
The proposed method is again the best or near-best on every gene, demonstrating
that robustness to process mismatch is achieved by construction rather than
by fortuitous cancellation of errors.

\begin{figure}[ht]
\centering
\includegraphics[width=1.0\textwidth]{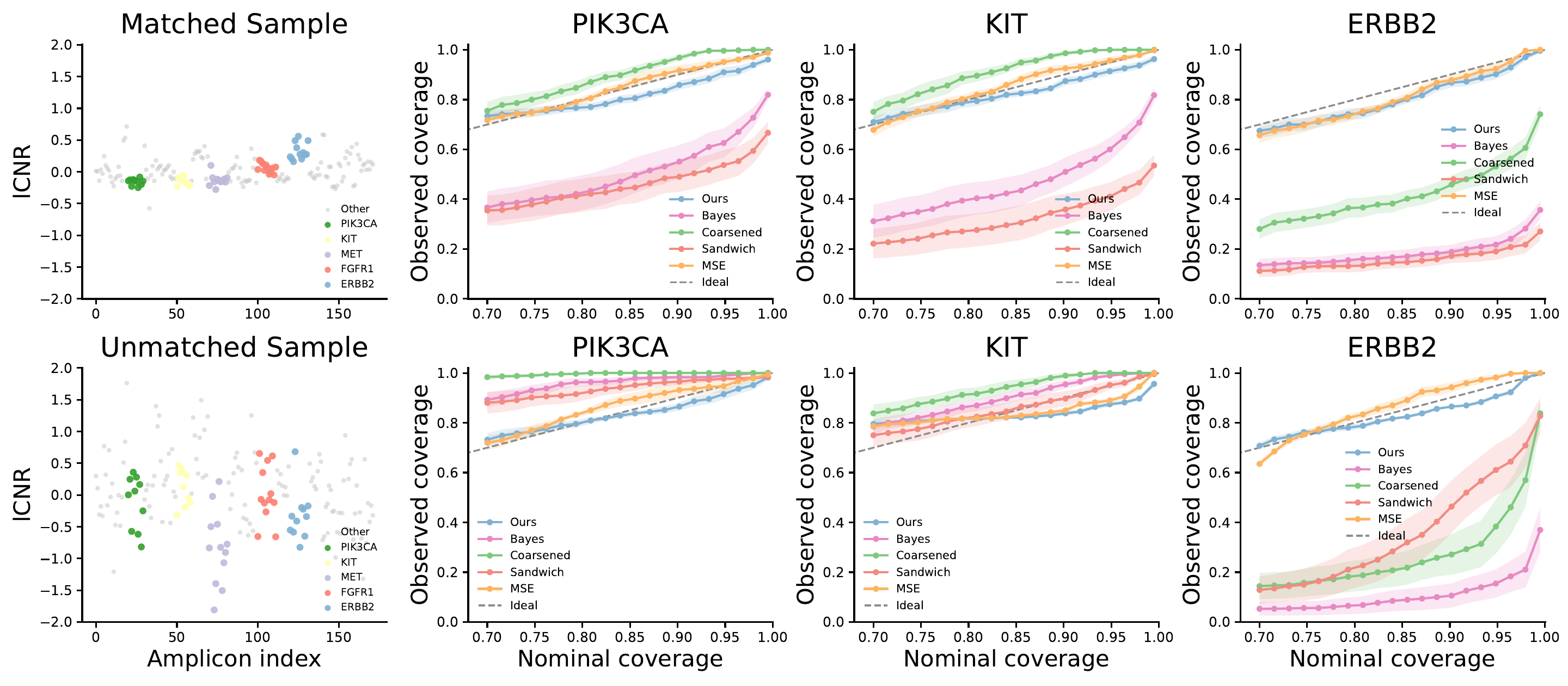}
\caption{Calibration curves of the various methods. \textbf{Top:} Coverages in the process-matched experiments. The far left plots the lCNR of a representative process-matched sample, with amplicons targeting the five CNV-relevant genes shown with different colors. The remaining plots show the true coverages on three of the five genes as a function of target coverages. The ideal is shown via a dotted line, while the average coverage and 95\% confidence interval for each of the five considered methods is shown in a different color\textbf{Bottom:} These plots have identical interpretation, except when the data come from unmatched (highly noisy) samples. }\label{fig:esscalibration}
\end{figure}

\begin{table}[htbp]
\centering
\caption{Mean absolute error $\times100$ with 95\% confidence intervals for coverage evaluation on the essential panel using matched and unmatched normals. Results are shown for five methods: Ours, MSE, Bayesian, Sandwich, and Coarsened}
\label{tab:essential_panel_coverage_mae}
\small
\setlength{\tabcolsep}{5pt}
\begin{tabular}{lcccccccccc}
\toprule
& \multicolumn{5}{c}{\textbf{Matched}} & \multicolumn{5}{c}{\textbf{Unmatched}} \\
\cmidrule(lr){2-6} \cmidrule(lr){7-11}
Method
& PIK3CA & KIT & MET & FGFR1 & ERBB2
& PIK3CA & KIT & MET & FGFR1 & ERBB2 \\
\midrule
Bayes & $33\pm6$ & $37\pm5$ & $40\pm7$ & $54\pm2$ & $66\pm3$ & $11\pm4$ & $12\pm1$ & $45\pm7$ & $25\pm6$ & $74\pm4$ \\
Coarsened & $6.6\pm.9$ & $8.0\pm1
$ & $18\pm5$ & $14\pm3$ & $42\pm3$ & $15\pm2$ & $9.5\pm2$ & $45\pm13$ & $10\pm2$ & $57\pm7$ \\
Sandwich & $38\pm6$ & $52\pm6$ & $47\pm7$ & $60\pm3$ & $69\pm3$ & $10\pm2$ & $7.6\pm1.4$ & $43\pm13$ & $20\pm4$ & $49\pm8$ \\
MSE & $4.2\pm.5$ & $4.2\pm.6$ & $6.4\pm.6$ & $4.2\pm.7$ & $5.6\pm.5$ & $5.7\pm.5$ & $4.4\pm.5$ & $8.1\pm2$ & $7.4\pm.5$ & $4.2\pm.7$ \\
Gamma & $4.9\pm.2$ & $4.4\pm.3$ & $6.1\pm.6$ & $5.3\pm.4$ & $3.0\pm.2$& $4.4\pm.6$ & $5.4\pm.1$ & $5.5\pm.7$ & $7.0\pm.8$ &  $3.2\pm.2$\\
\bottomrule
\end{tabular}
\end{table}

\subsubsection{Heterogeneous Mixture}
\label{ssec:essential_mixed}

\begin{figure}[ht]
\centering
\includegraphics[width=1.0\textwidth]{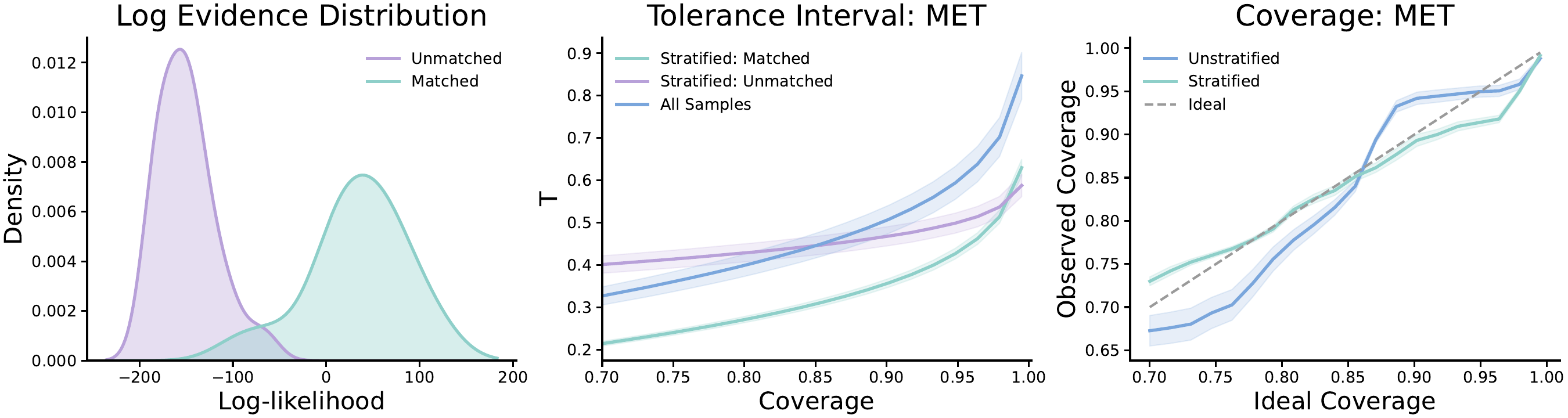}
\caption{The effect of stratefication in heterogeneous mixtures. \textbf{(A)} The distributions of log-likelihoods in the process matched and unmatched populations. \textbf{(B)} The tolerance intervals for MET on each of the subgroups for a stratefied fit, and all samples when group differences are ignored. \textbf{(C)} The associated coverages for each method.}\label{fig:heterogeneous}
\end{figure}

We then evaluate a situation that we frequently encounter, when both matched and unmatched samples are included in the same run. We evaluate this by combining the data from both the highest and lowest values of process matching in the experiment above, resulting in 30 trials of 39 samples, of which 17 are process matched and 22 are not process matched. This situation is evident when we examine the distribution of likelihoods, shown in panel (A) of Figure~\ref{fig:heterogeneous}. Unsurprisingly, the unmatched samples appear noisier and have correspondingly lower likelihoods. 

We compared the results of our method with and without stratification. Note that since 22 samples are from one group and 17 in the other, this analysis does not ``cheat'' by having the median perfectly align with group membership. We first show the performance guarantees for MET in panel (B) of Figure~\ref{fig:heterogeneous}. Note that the estimate does not lie between the estimates of the two groups over the entire interval; at the highest quantiles the estimate is excessively loose and larger than each group individually. This is the price we pay for modeling a mixture using a single parametric model. While we know that the quantile for the mixture is less than the maximum of the individual quantiles, with misspecified parametric models this is no longer true.  We can see in panel (C) of Figure~\ref{fig:heterogeneous} that the in the stratified model is better over the entire range, and importantly is similar to the aggregate guarantee, while having narrower intervals. Ironically, by modeling heterogeneity in the noise distribution, we are able to accurately state that our performance is better than our estimate when such heterogeneity is ignored.

\subsection{FFPE Validation Data}
\label{ssec:hs341}

\begin{figure}[ht]
\centering
\includegraphics[width=1.0\textwidth]{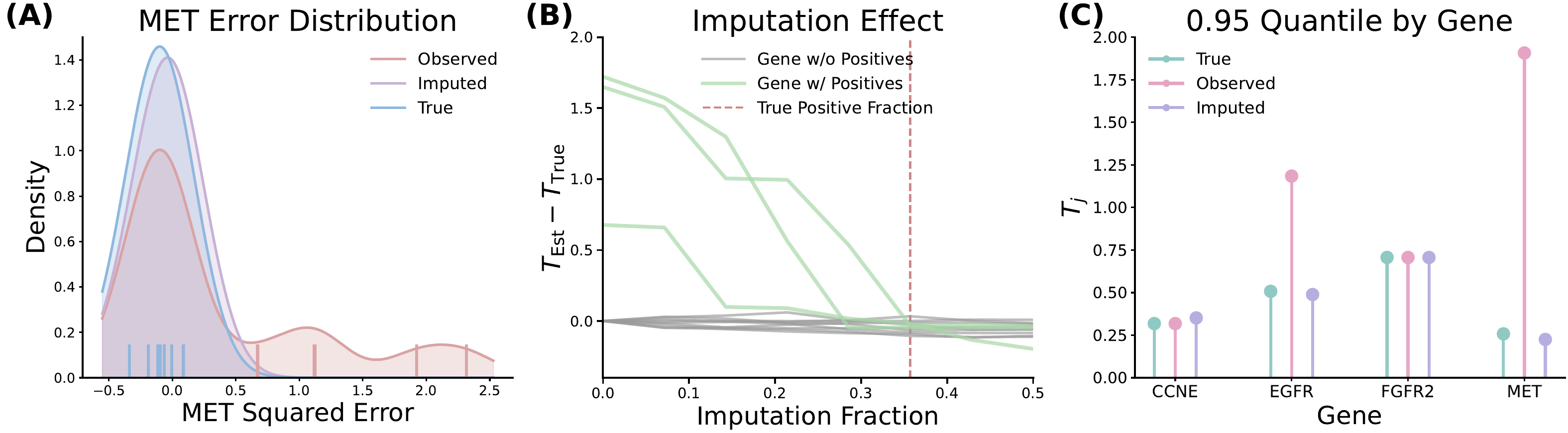}
\caption{The effect of imputation on MET tolerance interval estimates.
\textbf{(A)} plots the KDEs for true, observed, and imputed MET error values.
\textbf{(B)} plots between estimated and true 0.95 quantiles across imputation fractions, with the dashed line indicating the true positive fraction.
\textbf{(C)} plots the true, observed, and imputed 0.95 quantiles for CCNE, EGFR, FGFR2, and MET.}\label{fig:5p3}
\end{figure}

Our final evaluation is performed on data for a panel designed for FFPE solid tumor diagnostics. The panel has 341 amplicons, of which 240 target 14 genes for CNV detection. This particular experiment was run internally with 14 samples, with 5 NIST standard samples CNV positive for ERBB2 and 5  positive for EGFR and MET. The effect of these outliers on the distribution of the MET losses is shown in panel (A) of Figure~\ref{fig:5p3}. Running this high of fraction of samples with CNVs in the same gene is inadvisable, as it can affect the baseline and obscures true signal. However, this is a common occurrence because samples with known CNVs are hard to procure, and validation runs naturally use multiple positive samples to ensure our methods work. As a result we can use this dataset to show the importance of the imputation step in~\ref{ssec:impute}, and evaluate the sensitivity to the imputation fraction.

To do so, we first evaluated the 95\% quantile on the samples, excluding the samples with known positives on a per-gene basis, which we treat as the ground-truth value. We then fit our method on the full dataset with the positives, increasing the fraction of imputed values up to 50\% of the samples. The difference between the estimated and true value is plotted in panel (B) of Figure~\ref{fig:5p3}.

We found that our method was relatively robust to imputation fraction, with the non-CNV genes having an average relative difference of 8\% when 30\% of the values are imputed, with a maximum of 22\%. Even when 50\% of the values are imputed the average relative difference was 15\% with a maximum of 38\%. While this is naturally suboptimal compared to no imputation, it is dramatically preferable to ignoring the outlier effect of true positives. If this imputation step were not performed, it would erroneously suggest that the limit of detection for MET was 8.6, as opposed to the true and far more reasonable value of 1.5 as shown in panel (C) of Figure~\ref{fig:5p3}. 

\section{Discussion} \label{sec6}
This work demonstrates that providing laboratory-specific frequentist performance guarantees for a Bayesian diagnostic pipeline requires a fundamentally different inferential strategy than improving the posterior itself. Our experiments demonstrate that state-of-the-art robust Bayesian modifications, the coarsened posterior~\citep{miller2019robust} and the sandwich adjustment~\citep{muller2013risk} fail to achieve reliable frequentist coverage on the small-amplicon panels typical of clinical NGS, because calibration is a property of the repeated-sampling procedure, not of any individual posterior. The proposed framework addresses this gap by recasting the problem as tolerance-interval estimation on empirical loss distributions, incorporating regularization from previous laboratories to improve performance and reduced variance. That the simpler MSE-based estimator also enters the single-digit MACE regime reinforces the broader lesson: the critical design choice is the frequentist reframing itself, with the Gamma model providing a further edge through its theoretical grounding in the noncentral chi-squared structure of the losses and its accommodation of process-mismatch bias via the shape parameter (Appendix A). Beyond CNV detection, this hybrid strategy, Bayesian inference for individual-level uncertainty paired with frequentist calibration for population-level guarantees, is applicable to any ML-based clinical assay that requires quantifiable performance bounds. The evidence-stratification component further highlights a principle with wider relevance in healthcare ML: subgroup-specific performance reporting, much like disaggregated evaluation in algorithmic fairness, can expose heterogeneity that aggregate metrics conceal.

Several limitations point to natural avenues for future work. The evidence-based stratification relies on a binary median split of the log model evidence $Z_s$, a deliberately simple scheme that treats process mismatch as a dichotomy and requires $K > 20$ to provide adequate per-stratum sample sizes. This is perhaps the most accessible opportunity for improvement: replacing the hard partition with continuous sample weights or data-driven clustering (Talbot and Ke, 2026) would allow finer-grained accommodation of heterogeneity without halving the effective sample size per stratum. Similarly, the conditional order-statistic imputation assumes a fixed number of suspect positives $m$, targeting amplifications whose effect on the upper tail is unbounded; extending the scheme to symmetric imputation for deletions and developing adaptive procedures for selecting $m$, for instance via goodness-of-fit testing on the bulk distribution, would broaden applicability to panels targeting homozygous losses.

\bibliography{sample}

@article{povysil2017panelcn,
  title={panelcn. MOPS: Copy-number detection in targeted NGS panel data for clinical diagnostics},
  author={Povysil, Gundula and Tzika, Antigoni and Vogt, Julia and Haunschmid, Verena and Messiaen, Ludwine and Zschocke, Johannes and Klambauer, G{\"u}nter and Hochreiter, Sepp and Wimmer, Katharina},
  journal={Human mutation},
  volume={38},
  number={7},
  pages={889--897},
  year={2017},
  publisher={Wiley Online Library}
}

@article{boeva2014multi,
  title={Multi-factor data normalization enables the detection of copy number aberrations in amplicon sequencing data},
  author={Boeva, Valentina and Popova, Tatiana and Lienard, Maxime and Toffoli, Sebastien and Kamal, Maud and Le Tourneau, Christophe and Gentien, David and Servant, Nicolas and Gestraud, Pierre and Rio Frio, Thomas and others},
  journal={Bioinformatics},
  volume={30},
  number={24},
  pages={3443--3450},
  year={2014},
  publisher={Oxford University Press}
}

@article{talevich2016cnvkit,
  title={CNVkit: genome-wide copy number detection and visualization from targeted DNA sequencing},
  author={Talevich, Eric and Shain, A Hunter and Botton, Thomas and Bastian, Boris C},
  journal={PLoS computational biology},
  volume={12},
  number={4},
  pages={e1004873},
  year={2016},
  publisher={Public Library of Science San Francisco, CA USA}
}

@article{moreno2020evaluation,
  title={Evaluation of CNV detection tools for NGS panel data in genetic diagnostics},
  author={Moreno-Cabrera, Jos{\'e} Marcos and Del Valle, Jes{\'u}s and Castellanos, Elisabeth and Feliubadal{\'o}, Lidia and Pineda, Marta and Brunet, Joan and Serra, Eduard and Capell{\`a}, Gabriel and L{\'a}zaro, Conxi and Gel, Bernat},
  journal={European Journal of Human Genetics},
  volume={28},
  number={12},
  pages={1645--1655},
  year={2020},
  publisher={Springer International Publishing Cham}
}

@article{grunwald2017inconsistency,
  title={Inconsistency of Bayesian Inference for Misspecified Linear Models, and a Proposal for Repairing It},
  author={Gr{\"u}nwald, Peter and van Ommen, Thijs},
  journal={Bayesian Analysis},
  volume={12},
  number={4},
  pages={1069--1103},
  year={2017}
}

@article{plagnol2012robust,
  title={A robust model for read count data in exome sequencing experiments and implications for copy number variant calling},
  author={Plagnol, Vincent and Curtis, James and Epstein, Michael and Mok, Kin Y and Stebbings, Emma and Grigoriadou, Sofia and Wood, Nicholas W and Hambleton, Sophie and Burns, Siobhan O and Thrasher, Adrian J and others},
  journal={Bioinformatics},
  volume={28},
  number={21},
  pages={2747--2754},
  year={2012},
  publisher={Oxford University Press}
}

@article{muller2013risk,
  title={Risk of Bayesian inference in misspecified models, and the sandwich covariance matrix},
  author={M{\"u}ller, Ulrich K},
  journal={Econometrica},
  volume={81},
  number={5},
  pages={1805--1849},
  year={2013},
  publisher={Wiley Online Library}
}

@article{klambauer2012cn,
  title={cn. MOPS: mixture of Poissons for discovering copy number variations in next-generation sequencing data with a low false discovery rate},
  author={Klambauer, G{\"u}nter and Schwarzbauer, Karin and Mayr, Andreas and Clevert, Djork-Arne and Mitterecker, Andreas and Bodenhofer, Ulrich and Hochreiter, Sepp},
  journal={Nucleic acids research},
  volume={40},
  number={9},
  pages={e69--e69},
  year={2012},
  publisher={Oxford University Press}
}

@article{jiang2018codex2,
  title={CODEX2: full-spectrum copy number variation detection by high-throughput DNA sequencing},
  author={Jiang, Yuchao and Wang, Rujin and Urrutia, Eugene and Anastopoulos, Ioannis N and Nathanson, Katherine L and Zhang, Nancy R},
  journal={Genome biology},
  volume={19},
  number={1},
  pages={202},
  year={2018},
  publisher={Springer}
}

@article{talbot2026bayescnv,
  title={BayesCNV: A Bayesian Hierarchical Model for Sensitive and Specific Copy Number Estimation in Cell Free DNA},
  author={Talbot, Austin and Kotlar, Alex and Rishishwar, Lavanya and Conley, Andrew and Zhao, Mengyao and Yang, Nachen and Liu, Michael and Wang, Zhaohui and Polvino, Sean and Ke, Yue},
  journal={Diagnostics},
  volume={16},
  number={2},
  pages={280},
  year={2026},
  publisher={MDPI}
}

@article{li2012contra,
  title={CONTRA: copy number analysis for targeted resequencing},
  author={Li, Jason and Lupat, Richard and Amarasinghe, Kaushalya C and Thompson, Ella R and Doyle, Maria A and Ryland, Georgina L and Tothill, Richard W and Halgamuge, Saman K and Campbell, Ian G and Gorringe, Kylie L},
  journal={Bioinformatics},
  volume={28},
  number={10},
  pages={1307--1313},
  year={2012},
  publisher={Oxford University Press}
}

@article{talbot2025classifying,
  title={Classifying Copy Number Variations Using State Space Modeling of Targeted Sequencing Data: A Case Study in Thalassemia},
  author={Talbot, Austin and Kotlar, Alex and Rishishiwar, Lavanya and Ke, Yue},
  journal={arXiv preprint arXiv:2504.10338},
  year={2025}
}

@article{krumm2012copy,
  title={Copy number variation detection and genotyping from exome sequence data},
  author={Krumm, Niklas and Sudmant, Peter H and Ko, Arthur and O'Roak, Brian J and Malig, Maika and Coe, Bradley P and NHLBI Exome Sequencing Project and Quinlan, Aaron R and Nickerson, Deborah A and Eichler, Evan E},
  journal={Genome research},
  volume={22},
  number={8},
  pages={1525--1532},
  year={2012},
  publisher={Cold Spring Harbor Laboratory Press}
}

@article{fromer2012discovery,
  title={Discovery and statistical genotyping of copy-number variation from whole-exome sequencing depth},
  author={Fromer, Menachem and Moran, Jennifer L and Chambert, Kimberly and Banks, Eric and Bergen, Sarah E and Ruderfer, Douglas M and Handsaker, Robert E and McCarroll, Steven A and O’Donovan, Michael C and Owen, Michael J and others},
  journal={The American Journal of Human Genetics},
  volume={91},
  number={4},
  pages={597--607},
  year={2012},
  publisher={Elsevier}
}

@article{babadi2023gatk,
  title={GATK-gCNV enables the discovery of rare copy number variants from exome sequencing data},
  author={Babadi, Mehrtash and Fu, Jack M and Lee, Samuel K and Smirnov, Andrey N and Gauthier, Laura D and Walker, Mark and Benjamin, David I and Zhao, Xuefang and Karczewski, Konrad J and Wong, Isaac and others},
  journal={Nature genetics},
  volume={55},
  number={9},
  pages={1589--1597},
  year={2023},
  publisher={Nature Publishing Group US New York}
}

@article{miller2019robust,
  title={Robust Bayesian inference via coarsening},
  author={Miller, Jeffrey W and Dunson, David B},
  journal={Journal of the American Statistical Association},
  year={2019},
  publisher={Taylor \& Francis}
}

@article{bours2021bayes,
  title={Bayes’ rule in diagnosis},
  author={Bours, Martijn JL},
  journal={Journal of Clinical Epidemiology},
  volume={131},
  pages={158--160},
  year={2021},
  publisher={Elsevier}
}

@article{gelman2020bayesian,
  title={Bayesian analysis of tests with unknown specificity and sensitivity},
  author={Gelman, Andrew and Carpenter, Bob},
  journal={Journal of the Royal Statistical Society Series C: Applied Statistics},
  volume={69},
  number={5},
  pages={1269--1283},
  year={2020},
  publisher={Oxford University Press}
}

@inproceedings{lotfi2022bayesian,
  title={Bayesian model selection, the marginal likelihood, and generalization},
  author={Lotfi, Sanae and Izmailov, Pavel and Benton, Gregory and Goldblum, Micah and Wilson, Andrew Gordon},
  booktitle={International Conference on Machine Learning},
  pages={14223--14247},
  year={2022},
  organization={PMLR}
}

@article{lee2024using,
  title={Using Bayesian statistics in confirmatory clinical trials in the regulatory setting: a tutorial review},
  author={Lee, Se Yoon},
  journal={BMC Medical Research Methodology},
  volume={24},
  number={1},
  pages={110},
  year={2024},
  publisher={Springer}
}

@article{peng2015reducing,
  title={Reducing amplification artifacts in high multiplex amplicon sequencing by using molecular barcodes},
  author={Peng, Quan and Vijaya Satya, Ravi and Lewis, Marcus and Randad, Pranay and Wang, Yexun},
  journal={BMC genomics},
  volume={16},
  number={1},
  pages={589},
  year={2015},
  publisher={Springer}
}

@article{derouault2020covcopcan,
  title={CovCopCan: An efficient tool to detect Copy Number Variation from amplicon sequencing data in inherited diseases and cancer},
  author={Derouault, Paco and Chauzeix, Jasmine and Rizzo, David and Miressi, Federica and Magdelaine, Corinne and Bourthoumieu, Sylvie and Durand, Karine and Dzugan, Helene and Feuillard, Jean and Sturtz, Franck and others},
  journal={PLoS Computational Biology},
  volume={16},
  number={2},
  pages={e1007503},
  year={2020},
  publisher={Public Library of Science San Francisco, CA USA}
}

@article{gao2019next,
  title={Next Generation-Targeted Amplicon Sequencing (NG-TAS): an optimised protocol and computational pipeline for cost-effective profiling of circulating tumour DNA},
  author={Gao, Meiling and Callari, Maurizio and Beddowes, Emma and Sammut, Stephen-John and Grzelak, Marta and Biggs, Heather and Jones, Linda and Boumertit, Abdelhamid and Linn, Sabine C and Cortes, Javier and others},
  journal={Genome medicine},
  volume={11},
  number={1},
  pages={1},
  year={2019},
  publisher={Springer}
}

@article{sie2014performance,
  title={Performance of amplicon-based next generation DNA sequencing for diagnostic gene mutation profiling in oncopathology},
  author={Sie, Daoud and Snijders, Peter JF and Meijer, Gerrit A and Doeleman, Marije W and van Moorsel, Marinda IH and van Essen, Hendrik F and Eijk, Paul P and Gr{\"u}nberg, Katrien and van Grieken, Nicole CT and Thunnissen, Erik and others},
  journal={Cellular Oncology},
  volume={37},
  number={5},
  pages={353--361},
  year={2014},
  publisher={Springer}
}

@article{karapetis2008k,
  title={K-ras mutations and benefit from cetuximab in advanced colorectal cancer},
  author={Karapetis, Christos S and Khambata-Ford, Shirin and Jonker, Derek J and O'Callaghan, Chris J and Tu, Dongsheng and Tebbutt, Niall C and Simes, R John and Chalchal, Haji and Shapiro, Jeremy D and Robitaille, Sonia and others},
  journal={New England Journal of Medicine},
  volume={359},
  number={17},
  pages={1757--1765},
  year={2008},
  publisher={Mass Medical Soc}
}

@article{fehrenbacher2020nsabp,
  title={NSABP B-47/NRG oncology phase III randomized trial comparing adjuvant chemotherapy with or without trastuzumab in high-risk invasive breast cancer negative for HER2 by FISH and with IHC 1+ or 2+},
  author={Fehrenbacher, Louis and Cecchini, Reena S and Geyer Jr, Charles E and Rastogi, Priya and Costantino, Joseph P and Atkins, James N and Crown, John P and Polikoff, Jonathan and Boileau, Jean-Francois and Provencher, Louise and others},
  journal={Journal of clinical oncology},
  volume={38},
  number={5},
  pages={444--453},
  year={2020},
  publisher={American Society of Clinical Oncology}
}

@article{bissiri2016general,
  title={A general framework for updating belief distributions},
  author={Bissiri, Pier Giovanni and Holmes, Chris C and Walker, Stephen G},
  journal={Journal of the Royal Statistical Society Series B: Statistical Methodology},
  volume={78},
  number={5},
  pages={1103--1130},
  year={2016},
  publisher={Oxford University Press}
}

@article{romond2012seven,
  title={Seven-year follow-up assessment of cardiac function in NSABP B-31, a randomized trial comparing doxorubicin and cyclophosphamide followed by paclitaxel (ACP) with ACP plus trastuzumab as adjuvant therapy for patients with node-positive, human epidermal growth factor receptor 2--positive breast cancer},
  author={Romond, Edward H and Jeong, Jong-Hyeon and Rastogi, Priya and Swain, Sandra M and Geyer Jr, Charles E and Ewer, Michael S and Rathi, Vikas and Fehrenbacher, Louis and Brufsky, Adam and Azar, Catherine A and others},
  journal={Journal of Clinical Oncology},
  volume={30},
  number={31},
  pages={3792--3799},
  year={2012},
  publisher={American Society of Clinical Oncology}
}

@article{neal2001annealed,
  title={Annealed Importance Sampling},
  author={Neal, Radford M},
  journal={Statistics and Computing},
  volume={11},
  number={2},
  pages={125--139},
  year={2001},
  publisher={Springer}
}

@book{gelman1995bayesian,
  title={Bayesian Data Analysis},
  author={Gelman, Andrew and Carlin, John B and Stern, Hal S and Rubin, Donald B},
  year={1995},
  publisher={Chapman and Hall/CRC}
}

@article{urkowitz2005energy,
  title={Energy detection of unknown deterministic signals},
  journal={Proceedings of the IEEE},
  author={Urkowitz, Harry},
  volume={55},
  number={4},
  pages={523--531},
  year={1967},
  publisher={IEEE}
}

@book{bickel2015mathematical,
  title={Mathematical statistics: basic ideas and selected topics, volumes I-II package},
  author={Bickel, Peter J and Doksum, Kjell A},
  year={2015},
  publisher={Chapman and Hall/CRC}
}

@article{talbot2026detecting,
  title={Detecting Batch Heterogeneity via Likelihood Clustering},
  author={Talbot, Austin and Ke, Yue},
  journal={arXiv preprint arXiv:2601.09758},
  year={2026}
}

@book{david2004order,
  title={Order statistics},
  author={David, Herbert A and Nagaraja, Haikady N},
  year={2004},
  publisher={John Wiley \& Sons}
}

@article{fowler2016accurate,
  title={Accurate clinical detection of exon copy number variants in a targeted NGS panel using DECoN},
  author={Fowler, Anna and Mahamdallie, Shazia and Ruark, Elise and Seal, Sheila and Ramsay, Emma and Clarke, Matthew and Uddin, Imran and Wylie, Harriet and Strydom, Ann and Lunter, Gerton and others},
  journal={Wellcome open research},
  volume={1},
  pages={20},
  year={2016}
}

@article{munte2025detection,
  title={Detection of germline CNVs from gene panel data: benchmarking the state of the art},
  author={Munt{\'e}, Elisabet and Roca, Carla and Del Valle, Jes{\'u}s and Feliubadal{\'o}, Lidia and Pineda, Marta and Gel, Bernat and Castellanos, Elisabeth and Rivera, Barbara and Cordero, David and Moreno, V{\'\i}ctor and others},
  journal={Briefings in Bioinformatics},
  volume={26},
  number={1},
  pages={bbae645},
  year={2025},
  publisher={Oxford University Press}
}

@article{betancourt2017conceptual,
  title={A conceptual introduction to Hamiltonian Monte Carlo},
  author={Betancourt, Michael},
  journal={arXiv preprint arXiv:1701.02434},
  year={2017}
}

@article{kleijn2012bernstein,
  title={The Bernstein-Von-Mises theorem under misspecification},
  author={Kleijn, BJK and van der Vaart, AW},
  journal={Electronic Journal of Statistics},
  volume={6},
  pages={354--381},
  year={2012},
  publisher={Institute of Mathematical Statistics}
}

@article{bingham2019pyro,
  title={Pyro: Deep universal probabilistic programming},
  author={Bingham, Eli and Chen, Jonathan P and Jankowiak, Martin and Obermeyer, Fritz and Pradhan, Neeraj and Karaletsos, Theofanis and Singh, Rohit and Szerlip, Paul and Horsfall, Paul and Goodman, Noah D},
  journal={Journal of machine learning research},
  volume={20},
  number={28},
  pages={1--6},
  year={2019}
}

@article{holmes2017assigning,
  title={Assigning a value to a power likelihood in a general Bayesian model},
  author={Holmes, Chris C and Walker, Stephen G},
  journal={Biometrika},
  volume={104},
  number={2},
  pages={497--503},
  year={2017},
  publisher={Oxford University Press}
}

@book{HuberRonchetti2009RobustStatistics,
  author    = {Huber, Peter J. and Ronchetti, Elvezio M.},
  title     = {Robust Statistics},
  edition   = {2nd},
  year      = {2009},
  publisher = {Wiley},
  address   = {Hoboken, NJ},
  isbn      = {9780470129906}
}

@article{gelman1998simulating,
  title={Simulating normalizing constants: From importance sampling to bridge sampling to path sampling},
  author={Gelman, Andrew and Meng, Xiao-Li},
  journal={Statistical science},
  pages={163--185},
  year={1998},
  publisher={JSTOR}
}

\newpage
\appendix
\section{Derivational Details}
\label{app:derivations}

\subsection{Moment-Matching for the Gamma Approximation}
\label{app:gamma_moments}

Let $Y = \tau_j^2\,\chi^2_1(\lambda_j)$ where $\chi^2_1(\lambda_j)$ is
the noncentral chi-squared distribution with one degree of freedom and
non-centrality~$\lambda_j$.  Its first two moments are
\[
  \mathbb{E}\bigl[\chi^2_1(\lambda_j)\bigr] = 1 + \lambda_j,
  \qquad
  \operatorname{Var}\bigl[\chi^2_1(\lambda_j)\bigr] = 2(1+2\lambda_j).
\]
Scaling by $\tau_j^2$ gives
\[
  \mathbb{E}[Y] = \tau_j^2(1+\lambda_j),
  \qquad
  \operatorname{Var}[Y] = 2\tau_j^4(1+2\lambda_j).
\]
For the $\operatorname{Gamma}(a,b)$ distribution (shape~$a$, scale~$b$)
we have $\mathbb{E}[Y]=ab$ and $\operatorname{Var}[Y]=ab^2$.  Equating:
\begin{align}
  ab   &= \tau_j^2(1+\lambda_j),    \label{eq:app_mean} \\
  ab^2 &= 2\tau_j^4(1+2\lambda_j).  \label{eq:app_var}
\end{align}
Dividing \eqref{eq:app_var} by \eqref{eq:app_mean}:
\[
  b_j = \frac{2\tau_j^2(1+2\lambda_j)}{1+\lambda_j},
\]
and substituting back:
\[
  a_j = \frac{(1+\lambda_j)^2}{2(1+2\lambda_j)}.
\]
Because $(1+\lambda_j)^2 - (1+2\lambda_j) = \lambda_j^2 \geq 0$, it
follows that $a_j \geq \tfrac{1}{2}$, with equality if and only if
$\lambda_j = 0$ (no process-mismatch bias).  Fitting a Gamma by maximum
likelihood therefore anchors at $a_j = \tfrac{1}{2}$ in the unbiased
regime and absorbs the non-centrality into an inflated shape when bias
is present.

\subsection{Fisher Consistency Constant for the MAD}
\label{app:mad_constant}

For $X \sim \mathcal{N}(\mu,\,\sigma^2)$, the median absolute deviation
is $\operatorname{MAD}(X) = \sigma\,\Phi^{-1}(3/4)$, where~$\Phi$ is
the standard Normal CDF.  To make the MAD a consistent estimator of the
standard deviation~$\sigma$, one multiplies by
\[
  c_n = \frac{1}{\Phi^{-1}(3/4)} \approx 1.4826,
\]
so that $\hat\sigma = c_n \cdot \operatorname{MAD}$
satisfies $\mathbb{E}[\hat\sigma] = \sigma$ under
normality~\citep{HuberRonchetti2009RobustStatistics}.  The resulting
estimator retains the 50\% breakdown point of the median while providing
an asymptotically unbiased scale estimate.

\section{Effect of Poor Process Matching}
\label{app:process_matching}

\begin{figure}[ht]
\centering
\includegraphics[width=1.0\textwidth]{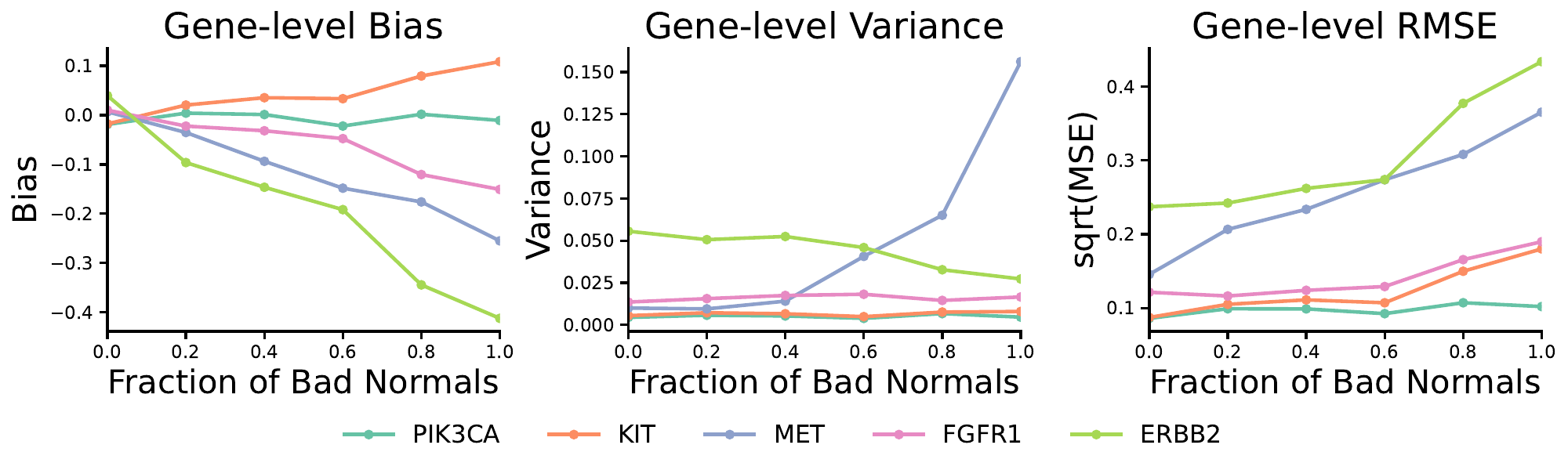}
\caption{The bias/variance decomposition of CNV estimates as a function of noraml matching}\label{appfig1}
\end{figure}

In this section we show the impact that a lack process matching can have on the CNV estimates. All of these 32 samples are normal in a 172 amplicon panel targeting 5 genes for CNV prediction. Of these 32 samples, 10 are substantially degraded. We estimate the CNVs in the genes using 5 samples as normal, ranging from all 5 coming from the good samples, to all 5 coming from the bad samples. We then bootstrap these estimates over different choices of normal and evaluate the posterior means in the remaining 17-22 good samples. The results are plotted in figure~\ref{appfig1}.

We can see that two genes are strongly affected by a lack of process matching, MET and ERBB2. However, they are affected differently, with ERBB2 suffering from dramatic biases while MET has high variance estimates. As a result, the assumption that a lCNR of 0 is not valid when samples are not process matched, and our confidence intervals must reflect this.

\end{document}